\documentclass[12pt]{article}

\usepackage{mathptmx}

\usepackage[dvipsnames]{xcolor}
\usepackage{physics}
\usepackage{braket}
\usepackage{scalerel}
\usepackage{blindtext} 
\usepackage{epsfig, cancel}

\usepackage{latexsym}
\usepackage[numbers,sort&compress]{natbib}
\usepackage{comment}
\usepackage{mathrsfs,amsmath,amssymb,amsthm,amsfonts,tikz,graphicx,accents,hyperref,color}
\usepackage{url}
\usepackage{dcolumn}
\usepackage{multirow}
\usepackage{color}
\usepackage{cancel}
\usepackage{soul}
\usepackage[normalem]{ulem}
\usepackage{txfonts}
\usepackage{epsfig}
\usepackage{psfrag}
\usepackage{subfigure}
\hypersetup{colorlinks=true}
\usepackage{mathtools}
\usepackage{enumitem}
\usepackage{float}
\usepackage{caption,ragged2e}
\usetikzlibrary{decorations.markings}

\usetikzlibrary{decorations.pathmorphing}

\def\H0{{\text{H}\hspace*{-2.05mm}\text{H} 0\hspace*{-1.35mm}0\ }}

\usepackage[all]{xy}

\usepackage{slashed}
\usepackage{slashed,ccaption}
\usepackage{multirow}
\usetikzlibrary{calc} 

\usepackage{caption}

\usepackage{array}
\newcommand{\Ottbar}{{\textrm{T}\hspace*{-1mm}{\textrm{T}}}}

\hypersetup{ linktoc=all,
    colorlinks, linkcolor={palatinateblue},
    citecolor={red}, urlcolor={amaranth} 
}

\graphicspath{{Images/}}

\renewcommand{\d}[1]{\ensuremath{\operatorname{d}\!{#1}}}

\DeclareSymbolFont{extraup}{U}{zavm}{m}{n}
\DeclareMathSymbol{\varheart}{\mathalpha}{extraup}{86}
\DeclareMathSymbol{\vardiamond}{\mathalpha}{extraup}{87}
\makeatletter
\renewcommand*{\@fnsymbol}[1]{\ensuremath{\ifcase#1\or \clubsuit \or \vardiamond \or \varheart\or
    \spadesuit\or \mathparagraph\or \|\or **\or \dagger\dagger
    \or \ddagger\ddagger \else\@ctrerr\fi}}
\makeatother

\definecolor{rosy}{RGB}{230,235,252}
\definecolor{myframetitle}{RGB}{90,89,170}
\definecolor{myblocktitle}{RGB}{140,185,249}
\definecolor{mytitle}{RGB}{10,80,26}

\definecolor{darkgreen}{RGB}{27,130,45}
\definecolor{darkblue}{rgb}{0,0,0.3}
\definecolor{darkred}{rgb}{0.7,0,0}

\definecolor{light gray}{RGB}{220,220,220}
\definecolor{dark purple}{RGB}{108,0,217}
\definecolor{pink}{RGB}{190,20,100}
\definecolor{orang}{RGB}{193,63,0}
\definecolor{green}{RGB}{11,98,17}
\definecolor{darkpink}{RGB}{153,0,76}
\definecolor{bluegreen}{RGB}{0,102,102}
\definecolor{greenlagan}{RGB}{0,102,0}
\definecolor{redgreen}{RGB}{102,102,0}
\definecolor{Redgreen}{RGB}{153,76,0}
\definecolor{vividviolet}{rgb}{0.62, 0.0, 1.0}
\definecolor{amaranth}{rgb}{0.9, 0.17, 0.31}
\definecolor{palatinateblue}{rgb}{0.15, 0.23, 0.89}
\definecolor{brightpink}{rgb}{1.0, 0.0, 0.5}
\definecolor{cornflowerblue}{rgb}{0.39, 0.58, 0.93}
\definecolor{deepcarminepink}{rgb}{0.94, 0.19, 0.22}
\definecolor{radicalred}{rgb}{1.0, 0.21, 0.37}

\newcommand\tcr{\textcolor{red}}

\newcommand\tcdg{\textcolor{darkgreen}}

\newcommand\ignore[1]{}
\usepackage[most]{tcolorbox}

\tcbset{highlight math style={left=02mm,right=02mm,top=02mm,bottom=02mm}} 
\usepackage{empheq}

\newcommand\inbox[1]{\tcbset{fonttitle=\scriptsize} \tcboxmath[colback=white,colframe=black!70]{#1}}

\begin{document}

\begin{titlepage}
\centering

{\Huge{\textbf{\tcr{GR from RG:}}}}\\ \vspace{2mm} {\large{ \textbf{\tcdg{Gravity Is Induced From Renormalization Group Flow In The Infrared}}}}

\vspace{0.5cm}

{\large M.~M.~Sheikh-Jabbari\footnote{Corresponding author. E-mail: jabbari@theory.ipm.ac.ir}$^{,a}$ and V.~Taghiloo\footnote{E-mail: v.taghiloo@iasbs.ac.ir}$^{,b,a}$}

\vspace{0.2cm}

\textit{$^a$ School of Physics, Institute for Research in Fundamental Sciences (IPM), \\ P.O.Box 19395-5531, Tehran, Iran}

\vspace{0.1cm}

\textit{$^b$ Department of Physics, Institute for Advanced Studies in Basic Sciences (IASBS), \\ P.O. Box 45137-66731, Zanjan, Iran}

\vspace{0.3cm}

{\large Date: \today}

\vspace{0.5cm}

\begin{center}
    \textbf{Abstract}
\end{center}
\begin{justify}
 In this essay and utilizing the holographic Renormalization Group (RG) flow, we demonstrate how the effective action of a non-gravitating quantum field theory in the ultraviolet (UV) develops an Einstein-Hilbert term in the infrared (IR). That is, gravity is induced by the RG flow. An inherent outcome of holography that plays a crucial role in our analysis is the  \textit{RG flow of boundary conditions}:  the rigid Dirichlet conditions on the background metric in the UV become an admixture of Dirichlet and Neumann as we flow to the IR, thereby ``unfreezing'' the metric and transforming it from a non-dynamical background into a dynamical field. This mechanism, which is a conceptually new addition to the standard Wilsonian RG flow, also provides the mechanism to evade the Weinberg-Witten no-go theorem. Within the GR from RG picture outlined here, the search for a quantum theory of gravity by treating the metric as a fundamental field may be a hunt for a phantom—akin to seeking the atomic structure of water by quantizing the equations of hydrodynamics.
\end{justify}

\vskip 1.5cm

\textit{
Received Honorable Mention in the 2026 Gravity Research Foundation Essay Competition.}

\end{titlepage}
\setcounter{page}{1}


\section{Quest for Quantum Gravity}

General Relativity (GR) is a remarkably successful classical theory of gravity, recasting the force as a manifestation of spacetime geometry and its dynamics. Its empirical triumphs encompass explaining the precession of Mercury’s orbit to the direct detection of gravitational waves and describing the cosmos at large scales. Quantum theory, on the other hand, provides an equally powerful description of the microscopic world and has been successfully tested to astounding levels of precision. Despite their individual successes, these two pillars of modern physics are profoundly incompatible. The quest for a consistent theory of ``Quantum Gravity'' (QGr) remains one of the greatest open problems in theoretical physics.

The first and most direct sign of conflict arises when one attempts to quantize GR as a field theory with the spacetime metric as a fundamental dynamical field. Doing so, the effective dimensionless coupling of gravity $G_N E^2$, grows with the square of energy, resulting in uncontrollable ultraviolet divergences that render the theory unpredictable at high energy (UV). This is not a mere technicality, but rather points to fundamental problems regarding the nature of spacetime and its quantum behavior. The need for a theory of QGr is not driven solely by this incompatibility. GR, despite its classical elegance and observational successes, suffers from theoretical issues like the existence of (curvature) singularities or closed timelike curves, which violate causality. Furthermore, the semi-classical description of black holes, yield black hole thermodynamics and evaporation of black holes via Hawking radiation, which in turn yields the challenging information paradox, see \cite{Grumiller:2022qhx} and references therein. 

Some different schools of thought have been developed to think about and hopefully to address the theoretical issues of GR and/or QGr; each pursuing the problem of QGr with its own philosophy and toolkit. The GR school, emphasizes on the geometric nature of GR, seeking to understand its quantum nature through methods like canonical quantization, the Wheeler-DeWitt equation, and modern loop quantum gravity; or try to replace differential geometry as the mathematical foundation of GR with a ``quantum spacetime'' (see e.g., \cite{Doplicher:1994tu}). The High-Energy Physics Theory (HEP-TH) school utilizes quantum field theory (QFT) as its primary toolkit and is dominated by string theory. Quantum Information school, is an emerging school of thought that tries to unite notions of spacetime and the fields defined on it via (quantum) information theoretic concepts. The borderline between these schools is not a bold one, and some frameworks like AdS/CFT \cite{Maldacena:1997re}, which we adopt in our analysis here, lie at the intersection of the three schools of thought mentioned above. See the first chapter of \cite{Grumiller:2022qhx} for more discussions and references.

Despite their vast differences in methodology, a common thread runs through many of these programs. They largely assume that gravity is a fundamental interaction of nature, a fundamental force. The central task, therefore, is to discover the correct way to quantize its fundamental degrees of freedom. There is, however, a complementary point of view or paradigm that gravity is an \textit{emergent}, non-fundamental force. This will be our main theme in this essay.

\paragraph{Emergent gravity paradigm.}
If gravity (GR) is an effective, collective, and coarse-grained description of an underlying microscopic theory that emerges in the long distances or low energies (IR), then the question of QGr becomes obsolete. There have been various attempts to formulate the emergent gravity paradigm, which we briefly review below. These formulations may be carried out within or inspired by either of the schools of thought mentioned above.

The emergent gravity ideas are supported by some different deductions within semiclassical GR. Black hole thermodynamics \cite{Bekenstein:1973ur, Hawking:1975vcx} hints to the existence of black hole microstates, which may be attributed to similar structures in the spacetime. The connection between gravity and thermodynamics seems to be deep and goes beyond black holes, as pointed out by Jacobson \cite{Jacobson:1995ab} and Padmanabhan \cite{Padmanabhan:2009vy}. Similar ideas has been resonated in the fluid/gravity correspondence \cite{Bhattacharyya:2007vjd}, Verlinde's entropic gravity \cite{Verlinde:2010hp}, and the ER=EPR conjecture \cite{Maldacena:2013xja} goes beyond the thermo/hydro -dynamics, suggesting that spacetime continuum and gravity has to do with information-theoretic structures like entanglement entropy and complexity \cite{Ryu:2006bv, VanRaamsdonk:2010pw, Susskind:2014rva}.

Accepting this paradigm brings a crucial shift in our central question. If the metric is a collective IR variable, then what are the fundamental gravity degrees of freedom (d.o.f.) in the deep UV? There are two conceivable options:
\begin{enumerate}
    \item The UV constituents are \textit{new, undiscovered degrees of freedom}, fundamentally gravitational in nature, from which spacetime emerges. String theory is a prime example here.
    \item The UV constituents are \textit{not new}, they are certain combinations of d.o.f. of a non-gravitational quantum field theory. In this view, gravity is a collective effect arising from the complex dynamics of ordinary quantum fields. 
\end{enumerate}
Our analysis and discussions in this essay point to the second option. In this framework, the graviton should be viewed as a collective excitation—analogous to a phonon in a crystal. Just as a phonon is a quantized mechanical wave that exists only as a collective motion of molecules in a crystal and has no meaning at scales smaller than the crystal lattice size, the graviton is a coherent, low-energy state of the true UV degrees of freedom, but it is absent in the fundamental UV spectrum. This viewpoint is by no means new. It was pioneered by Andrei Sakharov in 1967 \cite{Sakharov:1967pk}. Sakharov’s induced gravity proposed that the Einstein-Hilbert action could arise from the one-loop quantum fluctuations of standard matter fields. His method was to place a non-gravitational QFT on a fixed, classical, curved background and integrate out the high-energy modes. The resulting low-energy effective action indeed contained the Ricci scalar $R$ and higher-order curvature terms, suggesting that spacetime dynamics could be a byproduct of quantum fields. Sakharov's proposal suffers from a crucial setback: the metric tensor appearing in the effective action is that of the non-dynamical background, which was put in by hand; the metric was not allowed to fluctuate or respond to the matter sources. The challenge, then, is not merely to induce a geometric Einstein-Hilbert type term, but to explain how the geometry itself becomes a responsive, dynamical player.

The GR from RG proposal presented in \cite{Adami:2025pqr} that we review here, offers a solution to the above problem. Combining the machinery of holography with the logic of the Wilsonian Renormalization Group (RG), we demonstrate how a truly dynamical theory of gravity emerges in the IR from fundamentally non-gravitational modes in the UV; the holographic RG flow not only generates the gravitational action but simultaneously allows for the metric to become dynamical, rendering the metric from a static/fixed UV background into a dynamical field in the IR.

\section{GR From (Holographic) RG}

\begin{figure}[t]
\centering
\begin{tikzpicture}[scale=1.0]
    \coordinate (TL) at (-0.8,0);  \coordinate (TR) at (0.8,0);   
    \coordinate (BL) at (-0.8,-3); \coordinate (BR) at (0.8,-3);  

    \fill[gray!10] (-0.8,0) rectangle (0.8,-3);
    \fill[gray!20] (0,0) ellipse (0.8 and 0.2);
    \fill[gray!20] (0,-3) ellipse (0.8 and 0.2);

    \draw[darkred!60, thin, smooth] plot[variable=\y, domain=0:-3, samples=20] ({0.8*sin(-60) + 0.05*sin(180*\y)}, {\y + 0.1*cos(360*\y)});
    \draw[darkred!60, thin, smooth] plot[variable=\y, domain=0:-3, samples=20] ({0.8*sin(-30) - 0.04*sin(120*\y)}, {\y});
    \draw[darkred!60, thin, smooth] plot[variable=\y, domain=0:-3, samples=20] ({0.8*sin(0) + 0.06*sin(200*\y)}, {\y - 0.05*sin(360*\y)});
    \draw[darkred!60, thin, smooth] plot[variable=\y, domain=0:-3, samples=20] ({0.8*sin(30) + 0.03*sin(90*\y)}, {\y});
    \draw[darkred!60, thin, smooth] plot[variable=\y, domain=0:-3, samples=20] ({0.8*sin(60) - 0.05*cos(150*\y)}, {\y + 0.05*cos(180*\y)});

    \foreach \h in {-0.5, -1.0, -1.5, -2.0, -2.5} {
        \draw[darkred!60, thin, smooth] plot[variable=\t, domain=180:360, samples=20] ({0.8*cos(\t)}, {\h + 0.2*sin(\h*100)*sin(\t*3) + 0.05*sin(\t*5)});
    }

    \draw [darkred!80, very thick] (TL) -- (BL);
    \draw [darkred!80, very thick] (TR) -- (BR);
    \draw [darkred!80, very thick] (0,0) ellipse (0.8 and 0.2);
    \draw [darkred!80, very thick] (0,-3) ellipse (0.8 and 0.2);

    \draw [blue!60, thick] (0,0) ellipse (1.5 and 0.5);   
    \draw [blue!60, thick] (0,-3) ellipse (1.5 and 0.5);  
    \draw [blue!60, thick] (-1.5,0) -- (-1.5,-3);         
    \draw [blue!60, thick] (1.5,0) -- (1.5,-3);           

    \node [blue!50!black] at (0,-1.5) {$\mathcal{M}(r)$};
    \node [right, darkred!80, align=left, font=\footnotesize] (sigmalabel) at (0.71,-.85) {$\Sigma(r)$ };
    \node [blue!50!black] at (0,-1.5) {$\mathcal{M}(r)$};
    \node [right, blue, align=left, font=\footnotesize] (sigmalabel) at (1.5,-.1) {$\Sigma$ };
\end{tikzpicture}

\caption{\justifying{\footnotesize{Visualization of the emergence of dynamical gravity. The inner surface $\Sigma(r)$ represents the cutoff hypersurface at the renormalization scale $r$. The irregular, distorted grid is to stress the fact that the metric on $\Sigma(r)$, $\gamma_{ab}(r)$, is a dynamical fluctuating field. Its dynamics is governed by the RG flow of boundary conditions discussed in section \ref{sec:RG-bcs}. }}}
\label{fig:dynamical}
\end{figure}
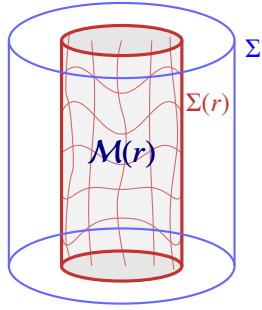

Our stage is a five-dimensional, asymptotically AdS spacetime, which we denote by $\mathcal{M}$. The geometry of this bulk spacetime is described by a theory of gravity, which, for simplicity, we take to be classical Einstein gravity. The AdS/CFT dictionary \cite{Maldacena:1997re, Witten:1998qj, Aharony:1999ti} posits that this 5D gravitational theory is dual to a 4D non-gravitational QFT, specifically a large $N$ conformal field theory, living on the timelike boundary of $\mathcal{M}$, denoted by $\Sigma$ and located at a radial coordinate $r \to \infty$. We foliate the bulk spacetime with a series of codimension-one hypersurfaces, $\Sigma(r)$, at constant values of the radial coordinate $r$, see FIG \ref{fig:dynamical}. The metric in the standard Fefferman-Graham gauge takes the form:
\begin{equation}
    ds^2 = g_{\mu\nu} dx^\mu dx^\nu = \frac{L^2}{r^2} dr^2 + h_{ab}(r, x) dx^a dx^b\, ,\qquad  h_{ab}:=\frac{r^2}{L^2}\gamma_{ab}\, ,
\end{equation}
where $L$ is the AdS$_5$ radius and $h_{ab}$ is the induced metric on the slice $\Sigma(r)$.

Two elements of the holographic dictionary crucial for our arguments are the physical meaning of the radial coordinate $r$ as the renormalization scale in the holographic dual QFT \cite{Susskind:1998dq, Peet:1998wn, deBoer:1999tgo, Skenderis:1999nb, Skenderis:2002wp} and that in the standard AdS/CFT dictionary one is prescribed to impose Dirichlet boundary conditions on the metric at the asymptotic boundary: $\delta h_{ab}(r \to \infty) = 0$ \cite{Witten:1998qj}. The asymptotic boundary at $r \to \infty$ corresponds to the deep UV of the field theory, while moving into the bulk to smaller $r$ corresponds to integrating out high-energy modes and flowing towards the IR. The radial evolution of physical quantities in the bulk therefore corresponds to the RG flow in the boundary QFT. The choice of Dirichlet boundary condition freezes the metric fluctuations at the boundary $\Sigma$, which physically means that by construction, the dual theory is a non-gravitational QFT in the UV.

The dynamics of the bulk is governed by the 5D Einstein-Hilbert action with a boundary term, evaluated over a finite region $\mathcal{M}(r_c)$ bounded by a cutoff surface $\Sigma({r_c})$:
\begin{equation}\label{unrenormalized-action}
    S_{r_c} = \frac{1}{2\kappa_5} \int_{\mathcal{M}(r_c)} \d{}^{5}x\, \sqrt{-g} \left( {R}[g] + \frac{12}{L^2} \right) + \frac{1}{\kappa_5} \int_{\Sigma({r_c})}\d{}^{4}x\,  \sqrt{-h}\, K\, ,
\end{equation}
where $\kappa_5$ is the 5D gravitational coupling. The resulting 5D Einstein equations can be decomposed into a set of evolution and constraint equations along the radial direction. On any such slice $\Sigma(r)$, we can define the Brown-York energy-momentum tensor (EMT) \cite{Brown:1992br}, $T_{ab}$, which is conjugate to the induced metric. The radial Hamiltonian constraint then takes the form of a crucial relation between the intrinsic curvature of the slice, ${R}[h]$, and a quadratic combination of this EMT \cite{Adami:2025pqr}:
\begin{equation}\label{Hamiltonian-constraint}
  \inbox{  {R}[h] + \frac{12}{L^2} + \kappa_5^2 \Ottbar = 0\,, \qquad  \Ottbar := T_{ab}T^{ab} - \frac{1}{3}T^2\,.}
\end{equation}

\paragraph{Computing the boundary effective action $S^{\text{bdy}}_{\Sigma(r)}$ at the RG scale $r/L^2$.} In the saddle-point approximation of holography, this boundary action is identified with the on-shell bulk gravity action \cite{Hartman:2018tkw, Parvizi:2025wsg}, $S^{\text{bdy}}_{\Sigma(r)} \equiv S_r[g]_{\text{on-shell}}$. A direct calculation shows that the changes in the on-shell action as we vary the cutoff $r$, is governed by the $\Ottbar$ operator \cite{McGough:2016lol, Taylor:2018xcy, Hartman:2018tkw, Parvizi:2025wsg}:
\begin{equation}
    r \frac{d}{dr} S^{\text{bdy}}_{\Sigma(r)} = -\frac{\kappa_5^2 L}{2} \int_{\Sigma(r)} \d{}^{4}x\,\sqrt{-h} \, \Ottbar\, .
\end{equation}
This is the RG flow equation for the boundary theory action and states that flowing from the UV towards the IR (decreasing $r$) deforms the boundary action by an operator that is quadratic in the energy-momentum tensor. This is a known feature of so-called $T\bar{T}$ deformations \cite{Smirnov:2016lqw}. 

The truly remarkable step, however, comes when we use the Hamiltonian constraint \eqref{Hamiltonian-constraint} to eliminate $\Ottbar$ in favor of purely geometric quantities and to find the remarkable result \cite{Adami:2025pqr}:
\begin{equation}\label{Sr-RG-flow}
    r \frac{d}{dr} S^{\text{bdy}}_{\Sigma(r)} = \frac{L}{2} \int_{\Sigma(r)} \d{}^{4}x\, \sqrt{-h} \left( {R}[h] + \frac{12}{L^2} \right)\, .
\end{equation}
This equation is at the heart of our analysis. Note that, while we used holography to derive it, both sides of this equation are only written in terms of quantities associated with $\Sigma(r)$, the dual QFT field and metric $h_{ab}$ defined on it. It demonstrates that the Wilsonian RG flow generates terms in the effective action that are precisely the Einstein-Hilbert action and a cosmological constant for the 4D metric $h_{ab}$. We began with a non-gravitational theory in the UV, but by integrating out high-energy modes, we have induced the building blocks of dynamical gravity. We stress that $r \frac{d}{dr} S^{\text{bdy}}_{\Sigma(r)}$ in \eqref{Sr-RG-flow} also involves the standard renormalization terms, like deformations by other irrelevant operators (than $\Ottbar$). However, since they are well-known and studied in the standard QFT textbooks, and to avoid cluttering, we have dropped those terms and only explicitly present the terms induced from $\Ottbar$ deformation. The contributions we are interested in are holographically captured by the $5d$ AdS-Einstein gravity \eqref{unrenormalized-action}.

The next step is to integrate this flow equation and determine the effective action of the dual QFT at the energy scale $\mu := r/L^2$. This is achieved by integrating \eqref{Sr-RG-flow} from an initial UV scale, associated with a large radius $r_0$, down to a lower energy scale $\mu$, corresponding to a radius $r<r_0$.
 Starting with a purely non-gravitational action in the UV, $S^{\text{bdy}}_{\Sigma_0} = S_{\text{QFT}}$, the integration yields the effective action for the boundary theory at the scale $\mu$, see \cite{Adami:2025pqr} for the details of the derivation:
\begin{equation}
    \inbox{\hspace*{-4mm}\begin{split}S^{\text{bdy}}_{\Sigma(\mu)} = S_{\text{QFT}} &+ \frac{1}{2\kappa_4(\mu)} \int_{\Sigma(\mu)} \d{}^{4}x\sqrt{-\gamma} \Big( {R}[\gamma] - 2\Lambda_4(\mu) \Big) + \beta(\mu) \int_{\Sigma(\mu)} \d{}^{4}x\,\sqrt{-\gamma} \, W_{abcd}W^{abcd} + \dots \hspace*{-4mm}\\ \kappa_4(\mu) &= \frac{2\kappa_5}{L^3} \left( \frac{1}{\mu^2} \right)\, , \qquad \Lambda_4(\mu) = 6\mu^2 \left( \frac{\mu_0^4 - \mu^4}{\mu_0^4} \right)\, , \qquad \beta(\mu) = \frac{\kappa_5}{4L^3} \ln \left( \frac{\mu}{\mu_0} \right)\, .
    \end{split}}
\end{equation}
Here, $\gamma_{ab}(r)$ is the conformal metric on the cutoff slice, and the ellipsis denotes higher-order corrections in the asymptotic expansion. 
This result is the central equation of our analysis, and its interpretation is profound. We began in the UV (large $\mu_0$) with a standard, non-gravitational quantum field theory, where the effective 4D Newton constant is zero ($\kappa_4 \to 0$), and there is no Einstein-Hilbert term in the action. As we coarse-grain our description and flow towards the IR (smaller $\mu$), the holographic RG flow yields/induces a full-fledged gravitational action. A 4D Newton's constant emerges, a cosmological constant is induced, and even higher-curvature terms like the Weyl-squared action appear with their own running couplings. Gravity, in this picture, is not a fundamental force that we must quantize; it is an effective, long-distance description that is inevitably induced by the quantum dynamics of an underlying non-gravitational system. The force that holds the cosmos together is an emergent phenomenon, a reinterpretation of deformations by the $\Ottbar$ operator induced by quantum effects of all the existing non-gravitating fields, as we zoom out to macroscopic scales.

\section{RG Flow of Boundary Conditions}\label{sec:RG-bcs}
The emergence of a gravitational action, as already mentioned in the review of Sakharov's idea \cite{Sakharov:1967pk}, does not mean we have a gravitating theory. For a true gravitating theory, the metric field $\gamma_{ab}$ must be allowed to fluctuate; it cannot remain a fixed background structure. This may pose a potential paradox: as a part of our AdS/CFT framework, we imposed the Dirichlet boundary condition $\delta \gamma_{ab}(\infty) = 0$ at the UV boundary of AdS, which turns gravity off. Then, how can the metric become dynamical in the IR? The answer lies in the fact that together with all the other things, the boundary conditions also have their own ``holographic RG flow''; the boundary condition on the metric $\gamma_{ab}$ at a finite $r$ does not remain Dirichlet \cite{deHaro:2000wj, Heemskerk:2010hk, Adami:2025pqr, Sheikh-Jabbari:2025kjd}.

The fields at the UV boundary, $\{\gamma_{ab}(\infty), T_{ab}(\infty)\}$, are not independent of the fields at a finite cutoff slice $\Sigma(r_c)$ with radius $r_c$, $\{\gamma_{ab}(r_c), T_{ab}(r_c)\}$. They are connected by the bulk equations of motion, which dictate the radial evolution between the two slices. In principle, one can solve these evolution equations to express the UV data as a functional of the IR data: 
\begin{equation}
    \gamma_{ab}(\infty) = \mathcal{G}_{ab}[\gamma_{ab}(r_c), T_{ab}(r_c)]\, , \quad T_{ab}(\infty) = \mathcal{T}_{ab}[\gamma_{ab}(r_c), T_{ab}(r_c)]\, .
\end{equation}
The original Dirichlet boundary condition $\delta \gamma_{ab}(\infty) = 0$ becomes a constraint on the variations of the IR fields: 
\begin{equation}
    \delta \mathcal{G}_{ab}[\gamma_{ab}(r_c), T_{ab}(r_c)] = 0\, .
\end{equation}
That is, a simple Dirichlet condition in the UV has morphed into what is generically a complicated, \textit{mixed} Dirichlet-Neumann boundary condition at the finite cutoff surface. It no longer demands that the metric be fixed, $\delta \gamma_{ab}(r_c) = 0$. Instead, it imposes a functional relationship between variations of the metric and variations of its conjugate momentum, the energy-momentum tensor.\footnote{The explicit asymptotic form is as follows \cite{Adami:2025pqr}, $
    \delta h_{ab}(\epsilon) = - \frac{\kappa_5}{5 L^3} \, \delta \Big( L^4 T_{ab}(\epsilon) \Big) + \mathcal{O}(\epsilon^2)\, , \text{with} \ \epsilon \equiv L^2/r^2 \ll 1$. For the AdS$_3$ example, where we have a much better control over the equations, the closed form (not as a large $r$ expanded version) has been worked out and discussed in \cite{Sheikh-Jabbari:2025kjd}. } This is precisely what is required for a dynamical theory of gravity. The metric is now free to fluctuate, as long as its fluctuations are correlated with those of the matter and energy sources described by $T_{ab}$. The RG flow of boundary conditions effectively ``unfreezes" the metric, turning it from a static background into a responsive, dynamical player.

Thus, the emergence of gravity is a two-fold process, perfectly captured by the holographic RG. The flow towards IR generates the Einstein-Hilbert action, providing the rules for the dynamics. Simultaneously, the flow of boundary conditions liberates the metric from its UV Dirichlet boundary condition prison, allowing it to participate in dynamics. The result is a consistent, self-contained picture where gravity is not postulated, but \textit{induced} by the RG flow. 

Therefore, the decades-long struggle to quantize GR by treating the metric as a fundamental field has been a struggle against a ghost. Such an approach is akin to searching for the atomic structure of water by mathematically quantizing the equations of hydrodynamics. In a fluid, quantization yields phonons, quasiparticles associated with the sound-waves that elegantly describe low-energy collective motion. Nonetheless, no amount of phonon physics can reveal the underlying molecules. Blindly pushing the quantization of the velocity field to higher energies leads only to mathematical breakdown and non-renormalizable divergences. These are not failures of the technique, but signals inapplicability of the fluid description. Our results suggest that the spacetime metric is the analogue of the fluid velocity field, and the graviton is the analogue of the phonons; \textit{GR is the hydrodynamics of the universe}. This view resonates with many similar statements as spin-offs of the AdS/CFT, e.g., see \cite{Polyakov:2006bz, Banks:2025erx}. The search for the atoms of spacetime within the spacetime geometry itself is a fruitless viewpoint; the graviton is a collective IR excitation and not a fundamental UV degree of freedom.

Our GR from RG construction identifies that the true UV constituents are \textit{not new} or exotic degrees of freedom. Gravity appears in the IR as a reinterpretation of the collective deformations of any QFT by its own $\Ottbar$ operator.
 
\section{Concluding Remarks}

\paragraph{Universality and background independence.} Einstein's gravity has two salient features: it is universal, and it is background independent. The former is made explicit by associating gravity with the fabric of the spacetime in which everything resides and in its general covariance, which is the invariance of the action under general coordinate transformations. The latter may be attributed to field redefinitions on the solution space of the theory \cite{Golshani:2024fry}. Our GR from RG program explains the universality of gravity: any matter in spacetime contributes to gravity, as by its mere existence, anything has an energy-momentum tensor and hence a non-zero $\Ottbar$ operator. From a different perspective, the universality of gravity implies that a similar underlying principle (general covariance) should work in any spacetime dimensions and should not be limited to the 4D case discussed here. In an upcoming work, we explicitly demonstrate how 2D gravity can be induced from the RG flow in a 2D (conformal) field theory \cite{GRfromRG_JT}. One may also observe that the background independence is also manifest. It is, nonetheless, instructive to discuss further and understand better the background independence within the GR from RG viewpoint.

\paragraph{Renormalized bulk theory.} The derivation reviewed thus far has relied on the unrenormalized bulk action \eqref{unrenormalized-action}. However, it is a well-known feature of the AdS/CFT correspondence that the on-shell bulk action diverges as one approaches the asymptotic boundary. To obtain finite, physically meaningful quantities, one must employ the procedure of holographic renormalization, adding covariant boundary counterterms to the action, see \cite{Skenderis:2002wp} and references therein. One can show that the GR from RG program can be extended to the holographically renormalized RG flows \cite{Adami:2025pqr}.

To this end, we may repeat the above analysis incorporating the standard counterterms required to render the variational principle well-posed \cite{Skenderis:2002wp}.  Performing the analyses as outlined above yields a renormalized induced action \cite{Adami:2025pqr}:
\begin{equation}
 \inbox{\begin{split}  S^{\text{ren-bdy}}_{\Sigma(\mu)} = & S^{\text{ren}}_{\text{QFT}} + \frac{1}{2\kappa_4^{\text{ren}}} \int_{\Sigma(\mu)} \d{}^{4}x \sqrt{-\gamma}\, {R}[\gamma] + \beta^{\text{ren}}(\mu) \int_{\Sigma(\mu)} \d{}^{4}x \sqrt{-\gamma} \, W_{abcd}W^{abcd} + \dots \\ &\Lambda_4^{\text{ren}} = 0\, , \qquad \kappa_4^{\text{ren}} = \mathrm{const}\, , \qquad \beta^{\text{ren}}(\mu) = \frac{L^3}{8 \kappa_5} \ln\left(\frac{\mu}{\mu_0}\right) - \frac{1}{4\kappa^{\text{ren}}_4} \left(\frac{1}{\mu} \right)^2\, . \end{split} }
\end{equation}
This result reveals two physically critical implications: (1) The effective cosmological constant $\Lambda_4^{\text{ren}}$ vanishes. Unlike the unrenormalized case where $\Lambda$ scales with the UV cutoff, here the divergences are systematically subtracted, suggesting a potential resolution to the cosmological constant problem: \textit{in a holographically induced gravity, the vacuum energy of the underlying QFT does not gravitate.} (2) The effective Newton constant $\kappa_4^{\text{ren}}$ becomes scale-independent. This stabilization removes the quadratic running characteristic of perturbative non-renormalizability in standard GR, yielding a finite gravitational coupling robust against standard UV divergence arguments.

\paragraph{How do we evade Weinberg-Witten No-Go Theorem?}

Any emergent gravity proposal must confront the formidable Weinberg-Witten no-go theorem \cite{Weinberg:1980kq}: A relativistic field theory cannot simultaneously possess a massless spin-2 particle (the graviton) and a local, Lorentz-invariant, conserved energy-momentum tensor (EMT) within the same Minkowski spacetime, $\partial_{a}T^{ab}=0$. Since standard non-gravitational QFTs naturally possess such an EMT, the theorem is widely interpreted as prohibiting the graviton to be a composite state of lower-spin particles. This appears to contradict our central idea, where in the UV we deal with a standard non-gravitating QFT.

The resolution to this apparent paradox is straightforward: the Weinberg-Witten theorem forbids gravity only in the presence of a conserved current $\partial_{a}T^{ab}=0$. In our framework, this condition holds strictly only at the asymptotic UV boundary, precisely where gravity is absent. As we flow to any finite energy scale $r$, the holographic RG forces the induced metric $\gamma_{ab}(r)$ to run, transforming the conservation law into the covariant form $\nabla_{a}T^{ab}=0$, which is a part of the 5D Einstein equations \cite{Adami:2025pqr}. The theorem does not preclude gravity with a \textit{covariantly conserved EMT}.\footnote{We note that in any diffeomorphism invariant field theory, we have a similar relation $\nabla_{a}T^{ab}=0$. This condition not only does not bar the presence of gravity, but  is crucial for the consistency of Einstein's equation.} Thus, our construction evades the no-go result; the 4D metric $\gamma_{ab}$ is a dynamical field and has energy scale dependence, therefore, the RG-induced gravity is compatible with the theorem.

\paragraph{Deviation from the standard Wilsonian RG.}
Standard textbook Wilsonian RG description asserts that couplings and dynamical fields should be renormalized while the background spacetime—typically flat Minkowski space—remains rigid and unrenormalized. Holography challenges this picture. Our analysis of the Hamiltonian constraint \eqref{Hamiltonian-constraint} makes this explicit: the textbook Wilsonian RG analysis ignores gravity. However, once 4D gravity is admitted, the $\Ottbar$ term in \eqref{Hamiltonian-constraint} acquires a non-vanishing coefficient, thereby coupling the background geometry to the stress-energy tensor. Since no symmetry protects $\Ottbar$, it is inevitably turned on by quantum effects as one flows to the IR. Consequently, the background metric cannot remain fixed and must evolve with the energy scale. This profound implication of holography—that the spacetime over which the 4D QFT resides, is subject to renormalization—calls for revisiting the Wilsonian paradigm and demands deeper investigation and understanding.
We close with the remark that, while we employed holography and 5D gravity to establish the emergence of gravity in the dual 4D theory, our analysis can, in principle, be carried out in a pure 4D picture with the modification in the Wilsonian RG discussed above. 

\section*{Acknowledgement}
We would like to thank Hamed Adami for his contribution to \cite{Adami:2025pqr}, upon which this work is based. We would like to thank Kostas Skenderis, Alessandro Tomasielo, and Mohammad Hassan Vahidinia for comments on the manuscript. We would also like to thank the members of the IPM HEP-TH weekly group meeting for insightful and useful discussions. MMShJ would like to thank BISMA, Beijing, for the hospitality during his visit in June-July 2025. The work of VT is supported by the Iran National Science Foundation (INSF) under project No. 4040771; MMShJ acknowledges the support within INSF research chair No. 40405163. 

\bibliographystyle{fullsort.bst}
\bibliography{reference}
\end{document}